\documentstyle[aasms4,12pt]{article}
\newcommand{\ts}{\thinspace}
\begin{document}

\title{$^{13}$CO($J$ = 1 -- 0) DEPRESSION IN LUMINOUS STARBURST MERGERS, REVISITED}

\author{Yoshiaki Taniguchi$^1$, Youichi Ohyama$^1$, \& D. B. Sanders$^2$}

\affil{$^1$Astronomical Institute, Tohoku University, Aoba, Sendai 980-8578, Japan}

\affil{$^2$Institute for Astronomy, University of Hawaii, 2680 Woodlawn Drive,
       Honolulu, HI 96822}


\begin{abstract}
It is known that merging galaxies with luminous starbursts and high far-infrared 
luminosities tend to have higher $R_{1-0} =^{12}$CO($J$=1--0)$/^{13}$CO($J$=1--0) 
integrated line intensity ratios
($R_{1-0} \simeq 20-50$) than normal spiral galaxies ($R_{1-0} \simeq 5-15$).
Comparing far-infrared luminosities [$L$(FIR)] with those of $^{12}$CO($J$=1--0) 
and $^{13}$CO($J$=1--0) for a sample of normal and starburst galaxies, 
Taniguchi \& Ohyama found that the observed high $R_{1-0}$ values for the 
luminous starburst mergers are attributed to their lower
$^{13}$CO line intensities by a factor of 3 on the average. 
They suggested the following two possibilities; in the luminous starburst mergers,
1) $^{13}$CO is underabundant with respect to $^{12}$CO, or 2) exitation and/or 
optical depth effects are responsible for the change in $R_{1-0}$.
In this paper, we investigate the second possibility using higher
transition data of both $^{12}$CO and $^{13}$CO emission lines.
Applying the same method proposed by Taniguchi \& Ohyama to both
$^{12}$CO($J$=2--1) and $^{13}$CO($J$=2--1), we find that $^{13}$CO($J$=2--1) 
is also depressed with respect to $^{12}$CO($J$=2--1).  This suggests that the 
$^{13}$CO gas may be underabundant in the high-$R_{1-0}$ starburst mergers 
although we cannot rule out the possibility that excitation and optical depth effects 
are still affecting $R_{2-1}$, for example due to the large velocity widths in the CO 
emission lines.  Additional observations of 
both $^{12}$CO and $^{13}$CO lines at $J \geq 3$ are required to better constrain 
the conditions of the molecular gas in luminous starburst galaxies.
\end{abstract}


\keywords{
galaxies: emission lines {\em -}
galaxies: starburst {\em -} interstellar: molecules}


\section{INTRODUCTION}

Since starburst phenomena affect both the chemical evolution of galaxies
and the physical conditions of the interstellar medium in galaxies, an understanding
of starburst activity is one of the important issues in astrophysics.
Starbursts occur in dense molecular gas clouds which are generally located 
in the central regions of galaxies. 
Massive stars formed in the starbursts provide a negative
feedback to the parent and ambient molecular gas clouds in the form of intense
radiation fields and strong stellar winds as well as subsequent supernova
explosions. Therefore it is important to investigate the molecular gas properties 
of starburst galaxies (Young \& Scoville 1991; Henkel, Mauersberger, \&
Baan 1991; Sanders \& Mirabel 1996 and references therein).

One of the interesting properties of molecular gas in starburst galaxies is 
that galaxy mergers with luminous starbursts (i.e., $L$(FIR){\ts}$\gtrsim 10^{11}L_{\odot}$: 
hereafter luminous starburst mergers) tend to have higher
$R_{1-0}$ ($\equiv I[^{12}$CO($J$=1--0)]/$I[^{13}$CO($J$=1--0)]) integrated line intensity ratios
than normal spiral galaxies (Aalto et al. 1991, 1995, 1997; Casoli et al. 1991;
Casoli, Dupraz, \& Combes 1992a, 1992b; Hurt \& Turner 1991;
Turner \& Hurt 1992; Garay, Mardones, \& Mirabel 1993;
Henkel \& Mauersberger 1993; Henkel et al. 1998).  One possible explanation for 
the higher $R_{1-0}$ values is that they are due to the inflow of disk gas 
with high $^{12}$C$/^{13}$C abundance ratios, possibly combined with a 
$^{12}$C enhancement caused by nucleosynthesis in massive stars (e.g., Henkel et al. 1998).
However, recently, Taniguchi \& Ohyama (1998a; hereafter TO98) have compared 
far-infrared luminosities [$L$(FIR)] with CO luminosities, $L[^{12}$CO($J$=1--0)$]$ and 
$L[^{13}$CO($J$=1--0)$]$, for a sample of normal and starburst galaxies, and have 
found that the observed higher $R_{1-0}$ values are associated  almost exclusively in 
the luminous starburst mergers and appear to be attributed to
a lower intensity of $^{13}$CO($J$=1--0) with respect to $^{12}$CO($J$=1--0) when 
compared to normal galaxies.  TO98 suggested either that $^{13}$CO is 
underabundant with respect to $^{12}$CO, or that the $^{13}$CO(1--0) level 
population is more depressed relative to $^{12}$CO(1--0) due to excitation 
and/or optical depth effects, leading to the high $R_{1-0}$ in the luminous 
starburst mergers studied in their paper. In this paper, we investigate the
second possibility using available higher transition data of both $^{12}$CO and $^{13}$CO 
emission lines.


\section{$^{12}$CO($J$=2--1)$/^{13}$CO($J$=2--1) INTEGRATED LINE INTENSITY RATIO}

TO98 demonstrated that the comparison of $L$(FIR) with both $L$[$^{12}$CO]
and $L$[$^{13}$CO] provides a powerful tool for understanding the origin 
of the high-$R_{1-0}$ in luminous starburst mergers (see also Taniguchi \& Ohyama 1998b).
If the observed high $R_{1-0}$ values are attributed to exication and optical depth effects, 
for example higher gas kinetic temperatures and/or denser gas clouds, then this could 
possibly be discerned in the measured values of $R_{2-1}$ or, if necessary, in even higher 
transition line ratios.   In order to examine if this is the case, 
we investigate the excitation properties of both $^{12}$CO and $^{13}$CO molecules
for a sample of normal and starburst galaxies using data currently available from the 
literature.

We have compiled $^{12}$CO($J$=2--1) and $^{13}$CO($J$=2--1) integrated intensities
from Aalto et al. (1995) and  Casoli et al. (1992b).
The integrated intensity ratio $I$[$^{12}$CO($J$=2--1)]$/I$[$^{13}$CO($J$=2--1)]
is referred as $R_{2-1}$.  Our sample consists of 24 galaxies and includes objects with 
extreme infrared luminosities such as the ultraluminous infrared galaxy Arp 220.
These integreated intensities are then used to compute CO luminosities;
$L$(CO) is defined as $L$(CO)$=A\times I$(CO) K km s$^{-1}$ pc$^2$
where $A$ is the observed area in units of pc$^2$ and
$I$(CO)$=\int T_{\rm A}^* \eta^{-1} dv$ K km s$^{-1}$
where $T_{\rm A}^*$ is the observed antenna temperature
corrected for atmospheric extinction and $\eta$ is the main beam efficiency.

The FIR data are compiled from the {\it IRAS} Faint Source Catalog
(Moshir et al. 1992).  The FIR luminosities are estimated using  $L$(FIR)$=4\pi D^2
1.26\times 10^{-11} [2.58\times S(60) + S(100)]$ (ergs s$^{-1}$) where 
$S$(60) and $S$(100) are the {\it IRAS} 60{\ts}$\mu$m and 100{\ts}$\mu$m fluxes
in units of Jy and $D$ is the distance (Helou, Soifer, \& Rowan-Robinson 1985).
Distances of nearby galaxies are taken from the Nearby Galaxies Catalog
(Tully 1988); distances of other galaxies are estimated using a Hubble constant
$H_0$ = 75 km s$^{-1}$ Mpc$^{-1}$ with $V_{\rm GSR}$ 
(recession velocity with respect to the Galactic Standard of Rest) given in
de Vaucouleurs et al. (1991).  The compiled data are given in Table 1.
All of the data presented here have been corected for beam-size 
(see Aalto et al. 1995; Casoli et al. 1992b).  Although our sample is not 
statistically complete, it is the largest sample compiled so far.
TO98 defined the class of high-$R_{1-0}$ galaxies by adopting the criterion of
$R_{1-0} \geq 20$. Using this limit, the present sample contains the following 
seven high-$R_{1-0}$ objects; NGC 1614, NGC 3256, NGC 4194, NGC 6240, 
Arp 220, Arp 299, and IRAS 18293$-$3413.

In Figure 1, we compare $L$[$^{12}$CO($J$=2--1)] with $L$[$^{13}$CO($J$=2--1)]. 
For reference, we also show the comparison between $L$[$^{12}$CO($J$=1--0)] 
and $L$[$^{13}$CO($J$=1--0)] in the left panel which is taken from TO98.
Although two high-$R_{1-0}$ galaxies, NGC 4194 and NGC 6240, have significantly 
lower $L$[$^{13}$CO($J$=2--1)] with respect to $L$[$^{12}$CO($J$=2--1)], 
the remaining galaxies have $R_{2-1}$ ratios within the upper range 
(i.e. $\sim$10--30) found for $R_{1-0}$.

In Figure 2, we compare $L$(FIR) with both $L$[$^{12}$CO($J$=2--1)] and 
$L$[$^{13}$CO($J$=2--1)].  $L$[$^{12}$CO($J$=2--1)] appears to be 
correlated with $L$(FIR) (in an integrated intensity versus flux plot, 
i.e. after removing the the $D^2$ effect from Figure 2, there is a good 
correlation), but the scatter is larger than observed for the  correlation 
between $L$[$^{12}$CO($J$=1--0)] and $L$(FIR) (TO98).  On the other hand, 
the correlation between $L$(FIR) and $L$[$^{13}$CO($J$=2--1)] is poorer 
than that between $L$(FIR) and $L$[$^{12}$CO($J$=2--1)] because the 
majority of the high-$R_{1-0}$ galaxies have lower $L$[$^{13}$CO($J$=2--1)] 
by a factor of 3 than what would be expected from the correlation for 
the normal-$R_{1-0}$ galaxies.  Thus, we find that both $^{12}$CO($J$=2--1) 
and $^{13}$CO($J$=2--1) show similar behavior as observed in the 
$J$=1--0 transition (TO98). In order to show this more clearly, we 
present a diagram of $R_{1-0}$ versus $R_{2-1}$ for our sample in Figure 3.
In Table 2, we summarize the statistical properties for both the high-$R_{1-0}$
and the ``normal-$R_{1-0}$" galaxies.

\section{$^{12}$CO($J$=3--2)$/^{13}$CO($J$=3--2) INTEGRATED LINE INTENSITY RATIO}

Published measurements of extragalactic $^{13}$CO($J$=3--2) emission are 
available only for M82 (Tilanus et al. 1991; Wild et al. 1992) and IC 342 
(Wall \& Jaffe 1990).  Tilanus et al. (1991) obtained $^{13}$CO($J$=3--2) 
spectra at three positions in M82; the center and the two peaks of the 
circumnuclear ring located $\pm$12 arcsec from the center using the 15{\ts}m 
JCMT with a beam size of 14 arcsec (FWHM). They obtained $R_{3-2} \simeq 15$ 
for the nucleus while $\simeq 10$ for the circumnuclear ring. On the other hand, 
Wild et al. (1992) measured both $^{12}$CO($J$=3--2) and $^{13}$CO($J$=3--2) emission 
lines using the IRAM 30{\ts}m radio telescope in February 1992 and obtained both
$I$[$^{12}$CO($J$=3--2)] = 1334 K km s$^{-1}$ at ($\Delta\alpha$, $\Delta\delta$)
= ($-$5, $-$5) and $I$[$^{13}$CO($J$=3--2)] = 70.4 K km s$^{-1}$ at 
($\Delta\alpha$, $\Delta\delta$) = ($-$7, $-$5) where $\Delta\alpha$ and 
$\Delta\delta$ are offsets from the nucleus position in right ascension and 
declination, respectively, in units of arcsec.  Although the measured positions 
are slightly different, these measurements give an integrated intensity ratio, 
$R_{3-2} \simeq 18.9$.  This together with the results by Tilanus et al. (1991) 
indicates that $R_{3-2}$ in the nuclear region is $\simeq$ 15 -- 20.
Since this value is nearly the same as the threshold value which defines 
the class of high-$R_{1-0}$ galaxies (TO98), it suggests that  
the $J = 3$ transition is still not high enough to allow dissentanglement 
of radiative transfer and abundance effects in M82.  

As suggested by 
Taniguchi \& Ohyama (1998b), it is 
possible that a large value of $R$ can be attributed to the effect of 
superwind activity (i.e., the possible destruction of dense gas as well 
as dust grains, and the large velocity widths observed in the CO outflow).  
In fact, in M82, $R_{3-2}$ is higher in the nuclear region than in the 
starburst ring (Tilanus et al. 1991).  Since M82 is indeed a superwind-starburst 
galaxy (Bland \& Tully 1988), the higher $R_{3-2}$ value in the nuclear 
region of M82 could possible be due to superwind activity.  However, Tilanus et al. 
(1991) suggested that intensity ratios of the three lowest transition lines 
of  $^{12}$CO and $^{13}$CO can be explained if $^{13}$CO  is overabundant 
with respect to $^{12}$CO just like what has been measured in the Galactic 
center, being contrary to our interpretation.

Another measurement of the $^{13}$CO($J$=3--2) line was obtained at the 
central region of IC 342; $R_{3-2} \simeq 7.7$  (Wall \& Jaffe 1990).
However, since the beam size of $^{13}$CO($J$=3--2) observation (24 arcsec) 
is different from that of $^{12}$CO($J$=3--2) one (15 arcsec), the above value 
may not be reliable.

Until more extragalactic CO($J$=3 -- 2) data is obtained it is 
clearly impossible to draw any firm conclusions about $R_{3-2}$ in 
luminous starburst mergers as well as normal galaxies. 

\section{DISCUSSION}

We have shown that the $R_{2-1}$ ratio is also high (typically $> 20$) in 
the high-$R_{1-0}$ galaxies. Furthermore, the $R_{3-2}$ value
in the nuclear region of M82 suggests that this ratio may also be high in 
luminous starbursts, but this is only for one object.
However, it is noted that the $^{12}$CO($J$=3--2)/$^{12}$CO($J$=1--0) 
integrated intensity ratio of 
starburst galaxies is often found to be higher than that in normal galaxies
(e.g., Devereux et al. 1994 and references therein).
Further, some nearby starburst galaxies such as M82 are
detected in CO($J$=4--3) (G\"usten et al. 1993) and in CO($J$=6--5)
(Harris et al. 1992). The detection of these higher-transition CO lines
suggests the presence of warm and dense gas clouds in starburst galaxies.

Although the kinetic gas temperature is not necessarily comparable 
to the dust temperature, it is interesting to compare molecular gas
properties with the dust temperature which is measured from the {\it IRAS} 
60 $\mu$m to 100 $\mu$m flux ratio, $S(60)/S(100)$.
In Figure 4, we compare $R_{2-1}$ with  $S(60)/S(100)$ for the galaxies
studied here. We also compare $R_{1-0}$ with $S(60)/S(100)$  
for reference (the left panel).
Though no tight correlation can be seen in either of these diagrams,
we find that both $R_{1-0}$  and $R_{2-1}$ tend to increase
with increasing $S(60)/S(100)$.
This tendency suggests that the galaxies with higher dust temperatures
have higher $R$ values on the average. 
In the lower panels of Figure 4, we show comparisons of $S(60)/S(100)$
with both $L$[$^{12}$CO($J$=2--1)]$/L$[$^{12}$CO($J$=1--0)] and 
$L$[$^{13}$CO($J$=2--1)]$/$L$[^{13}$CO($J$=1--0)] ratios.
Interestingly, we find no correlation in both the diagrams.
Average ratios of both $L$[$^{12}$CO($J$=2--1)]$/L$[$^{12}$CO($J$=1--0)] and
$L$[$^{13}$CO($J$=2--1)]$/L$[$^{13}$CO($J$=1--0)] are $0.81 \pm 0.17$ and 
$1.15 \pm 0.71$, respectively for the high $R_{1-0}$ starburst mergers. 
Therefore, there seems to be no significant difference in the excitation 
toward $J$=2 between $^{12}$CO and $^{13}$CO.

We also investigate whether a correlation exists between $R_{1-0}$ and 
the luminosity ratio $L$[$^{12}$CO($J$=2--1)]$/L$[$^{12}$CO($J$=1--0)] 
for our sample (Figure 5). Since the high-$R_{1-0}$ mergers tend to have 
higher dust temperatures (see Table 2), it is likely that their CO
kinetic temperatures are also higher than those of the normal-$R_{1-0}$
galaxies. These high temperatures could lead to both higher $R_{1-0}$
and to higher $L$[$^{12}$CO($J$=2--1)]$/L$[$^{12}$CO($J$=1--0)] 
as demonstrated by large velocity gradient models (Sakamoto et al. 
1994, 1997). However, there is no such tendency as shown in Figure 5.

Finally, we investigate whether there is any relationship between  
$R_{1-0}$, $R_{2-1}$ and the $L$(FIR)/$L$(CO) ratio which is 
generally considered to give a measure of star formation efficiency.
Figure 6 shows that the high-$R_{1-0}$  mergers as a group have a 
higher mean $L$(FIR)/$L$(CO) ratio as well as a higher mean $R_{2-1}$ 
ratio than the ``normal-$R_{1-0}$" galaxies.  But, other than that 
there appears to be no clear correlation between either $R_{1-0}$ or 
$R_{2-1}$ with respect to $L$(FIR)/$L$(CO).  Therefore, we are led 
once again to suggest that ``superwinds" may be the best explanation 
for what produces the high-$R_{1-0}$ values.  Indeed, it should be 
noted that nearly all of the high-$R_{1-0}$ galaxies show morphological 
and/or spectroscopic evidence for superwinds (Taniguchi \& Ohyama 1998b).  
Both the abnormally large velocity gradients assocated with these 
superwinds and the possible destruction of dense gas clouds by the 
dynamical effect of the superwind activity, could possibly  combine 
to reduce the observed intensity of the much more optically thin 
$^{13}$CO in the lower-$J$ transitions relative to the $^{12}$CO 
emission.  Further tests of this hypothesis will require measurements 
of the $J \geq 3$ transitions of both $^{12}$CO and $^{13}$CO for 
the galaxies in Table 1.

\vspace{1ex}

We would like to thank Seiichi Sakamoto for useful discussions.
YO was supported by a Grant-in-Aid for JSPS Fellows by
the Ministry of Education, Science, Sports and Culture.
This work was supported in part by the Ministry of Education, Science,
Sports and Culture in Japan under Grant Nos. 07055044, 10044052, and 10304013.


\newpage

\figcaption{
Diagram of $L$[$^{12}$CO($J$=2--1)] vs. $L$[$^{13}$CO($J$=2--1)] (right panel).
For reference, we also show a diagram  of $L$[$^{12}$CO($J$=1--0)] vs.
$L$[$^{13}$CO($J$=1--0)] taken from TO98 (left panel).
The high-$R$ galaxies ($R_{1-0} \geq 20$) are shown by filled circles. 
\label{fig1}
}

\figcaption{
Diagrams of $R_{2-1}$ (top), $L$[$^{12}$CO($J$=2--1)] (middle),
and $L$[$^{13}$CO($J$=2--1)] (bottom) against $L$(FIR).
The symbols have the same meaning as those in Figure 1.
Alphabets in the upper panel mean; a: NGC 6240, b: NGC 4194,
c: Arp 299, d: NGC 3256, e: Arp 220, and f: IRAS 18293$-$3413.
\label{fig2}
}

\figcaption{
Diagram of $R_{2-1}$ vs. $R_{1-0}$.
The symbols have the same meaning as those in Figure 1.
\label{fig3}
}

\figcaption{
Diagrams of $R_{2-1}$ (upper right), $R_{1-0}$ (upper left),
$^{12}$CO($J$=2--1) to $^{12}$CO($J$=1--0) luminosity ratio (lower right),
and $^{13}$CO($J$=2--1) to $^{13}$CO($J$=1--0) luminosity ratio  (lower left)
against $S(60)/S(100)$. Note that $R_{2-1}$ = 140 for NGC 6240.
The data of $R_{1-0}$  are taken from TO98. 
Alphabets in the upper left panel mean;
a: NGC 4194, b: NGC 6240, c: NGC 3256, d: NGC 3256, e: NGC 1614, f: NGC 3256,
g: IRAS 18293$-$3413, h: NGC 3256, i: Arp 220, j: Arp 299, k: Arp 299,
l: IRAS 18293$-$3413, m: ESO 541--IG 23, and  n: Arp 220. 
\label{fig4}
}

\figcaption{
Diagram of $R_{1-0}$ vs. $^{12}$CO($J$=2--1) to $^{12}$CO($J$=1--0)
integrated intensity ratio.
The symbols have the same meaning as those in Figure 1.
The large velocity gradient models taken from Sakamoto et al. (1997)
are show for the case of $X$(CO)$/dv/dr = 10^{-5}$ [km s$^{-1}$ pc$^{-1}$]$^{-1}$ 
where $X$(CO) is the fractional abundance of $^{12}$CO and $dv/dr$ is 
the velocity gradient. Kinetic temperatures and  densities of molecular gas
are labeled. 
\label{fig5}
}

\figcaption{
Diagrams of $R_{1-0}$ and $R_{2-1}$ against $L$(FIR)/$L$(CO) ratio.
The symbols have the same meaning as those in Figure 1.  Note that the data shown 
in the left panel are taken from TO98.
\label{fig6}
}

\end{document}